# Ab-initio investigation of Er$^{3+}$ defects in tungsten disulfide

Gabriel I. López-Morales,[1,2,3] Alexander Hampel,[4] Gustavo E. López,[2,3] Vinod M. Menon,[1,3] Johannes Flick,[4] Carlos A. Meriles[1,3*]

[1]*Department of Physics, City College of the City University of New York, New York, NY 10031 USA*
[2]*Department of Chemistry, Lehman College of the City University of New York, Bronx, NY 10468, USA*
[3]*The Graduate Center of the City University of New York, New York, NY 10016, USA*
[4] *Center for Computational Quantum Physics, Flatiron Institute, New York, NY 10010, USA*

We use density functional theory (DFT) to explore the physical properties of an Er$_W$ point defect in monolayer WS$_2$. Our calculations indicate that electrons localize at the dangling bonds associated with a tungsten vacancy (V$_W$) and at the Er$^{3+}$ ion site, even in the presence of a net negative charge in the supercell. The system features a set of intra-gap defect states, some of which are reminiscent of those present in isolated Er$^{3+}$ ions. In both instances, the level of hybridization is low, i.e., orbitals show either strong Er or W character. Through the calculation of the absorption spectrum as a function of wavelength, we identify a broad set of transitions, including one possibly consistent with the Er$^{3+}$ $^4I_{15/2} \rightarrow {^4I_{13/2}}$ observed in other hosts. Combined with the low native concentration of spin-active nuclei as well as the two-dimensional nature of the host, these properties reveal Er:WS$_2$ as a potential platform for realizing spin qubits that can be subsequently integrated with other nanoscale optoelectronic devices.

KEYWORDS: rare-earth ions, point defects, tungsten disulfide, optoelectronic properties, density functional theory

## I. INTRODUCTION

Single photon emitters in the form of rare-earth (RE) impurities in garnet minerals are being actively explored as solid-state spin-qubits, largely due to their electronically screened 4*f* states that result in hours-long coherence lifetimes[1] and, as in the case of Er$^{3+}$, homogeneous linewidths as narrow as 50 Hz in the telecom band.[2,3] In addition, these systems show promise for nanoscale sensing and as long-term quantum memories, with storage times exceeding a second.[4–7]

Despite the intrinsic potential of RE ions, most host materials bring in unwanted interactions, particularly in the form of deleterious hyperfine couplings between the ion and surrounding spin-active nuclei and/or impurities.[8,9] Dynamic decoupling techniques have been adapted to mitigate this problem, though at the expense of an increased complexity in the control protocols.[1,8] Isotopic depletion of spin-active nuclei can arguably prove effective but this route is not viable for all atomic species (e.g., yttrium), and material growth tends to be costly, particularly if the host contains heavy elements.[10,11] Even in the case of materials where hyperfine couplings are less critical (such as Er$^{3+}$:YSO or Nd$^{3+}$:YVO$_4$[3,8,12]) the problem of interfacing RE spins with optoelectronic elements remains a challenge as these hosts are not necessarily compatible with present material processing methods. All in all, these obstacles impose serious constraints for quantum information processing applications, particularly if scalability and device-integration requirements are taken into account.

With these considerations in mind, this work investigates the point defect formed by a substitutional Er impurity in tungsten disulfide (WS$_2$). Guiding this selection is the notion of the RE ion as a 'building block', rather insensitive to the host material of choice.[13] The system also leverages the two-dimensional (2D) nature of WS$_2$, suitable for integration into heterostructures and optoelectronic devices.[14–16] From among other candidate materials, tungsten disulfide is especially attractive for three main reasons: (1) the lattice is sufficiently large to accommodate a substitutional Er (taking the place of a W site) without introducing much strain, (2) a sizable bandgap of ~2.1 eV[16] makes it suitable for excitation and detection of photon emission in the near infrared, and (3) both S and W isotopes are mostly spin-less nuclei (their spin-active isotopes have natural abundances of only 0.7% and 14%, respectively) which results in an intrinsically low nuclear-spin material.

Recent years have seen various studies on rare-earth defects in solid-state systems using electronic structure methods. Defect systems require the explicit consideration of large system sizes (supercells) to accurately describe the host-defect interaction and limit defect-defect interactions. At the same time, the strongly correlated nature of the localized 4*f* states in $Er^{3+}$ limits the number of applicable electronic structure methods. While the many-body character of the $Er^{3+}$ could be described accurately by recent developments in so-called embedding techniques,[17–20] such calculations are computationally very demanding if large supercells are required. Due to its much more favorable scaling properties, in the past regular density-functional theory (DFT) methods, in particular the DFT+$U$ approach[21] has been used to characterize structural and optical properties of Er ions in GaN,[22,23] AlN,[24] and ZnO[25,26] hosts as well as other rare-earth ions such as Eu ions in GaN[27] with DFT+$U$ or hybrid-functional DFT.[28]

---

*email: cmeriles@ccny.cuny.edu

Following these approaches, in this work we use the DFT+$U$ approach to characterize the electronic properties of Er-ions in $WS_2$. Specifically, we determine the formation energies for this defect, and find it is most stable in two different charge states, namely, neutral, and negatively charged. In both instances, electrons localize at dangling bonds leaving behind an $Er^{3+}$ core. Our calculations indicate that the ground state of this system populates the highest energy levels in the $WS_2$ valence band; we also identify unoccupied intra-gap states reminiscent of the structure seen for the $Er^{3+}$ ion in other hosts. Solving the Casida equation and analyzing the resulting oscillator strengths, we conclude that a substitutional Er in $WS_2$ is potentially optically active near ~0.80 eV (1550 nm). Adding to ongoing work targeting spin-active, optically addressable photon emitters in 2D materials,[29–32] our results highlight the Er impurity in $WS_2$ as an intriguing material platform for applications in solid-state quantum technologies.

## II. COMPUTATIONAL METHODS

Our DFT calculations are performed within the projector-augmented wave (PAW) method using the Vienna *ab initio* Simulation Package (VASP).[33,34] All calculations are within the generalized-gradient approximation (GGA), using Perdew-Burke-Ernzerhof (PBE) functionals[35] to account for the exchange-correlation interaction. The atomic supercells are composed of 7 by 7 unit-cells for both pristine and Er-doped $WS_2$ monolayers, with a vacuum region of 20 Å along the z-axis to avoid spurious interactions between adjacent layers. Given the large supercell, and because the introduced point defect should yield localized atomic-like electronic states, it is reasonable to sample the irreducible Brillouin zone (IBZ) by using the $\Gamma$-point only. A high 700-eV plane-wave kinetic energy cut-off is chosen for all calculations. Convergence criteria for the electronic and the ionic loops are set to $10^{-8}$ eV and $10^{-4}$ eV/Å, respectively, to ensure convergence to the ground-state structure. Additionally (whenever specified), we have performed non-collinear calculations to add spin-orbit coupling (SOC) self-consistently. We note that in such calculations it is usually safer to switch off all symmetries, which translates in assuming that all atomic configurations belong to the $C_1$ point-group. For comparison, we have also employed calculations considering symmetry. In both cases, we find a lattice constant of 3.18 Å for unit-cell pristine $WS_2$, and direct bandgaps (at $K$) of 1.82 and 1.56 eV for collinear and SOC cases, respectively. These results agree well with previous calculations based on DFT within PBE.[36,37]

We characterize the overall electronic properties of Er-doped $WS_2$ monolayers by calculating the electronic density of states (DOS) using an 8 by 8 k-point grid, while the Kohn-Sham (KS) eigenvalues are calculated along high symmetry paths ($\Gamma$-M-K-$\Gamma$) to obtain the corresponding band structures. Additionally, the optical absorption coefficients are calculated using two different methods: (1) sum over all empty states within the independent particle approximation (IPA),[38] and (2) the Casida equation (within the Tamm-Dancoff approximation (TDA))[39–41] for several valence/conduction states, both at the $\Gamma$-point. We perform both sets of calculations at the DFT level, namely, using the PBE exchange-correlation functional.

## III. RESULTS AND DISCUSSION

Figure 1(a) shows top and angle views of a substitutional Er impurity in a $WS_2$ monolayer for the negatively charged state. Relaxing the defect structure and comparing it to the bulk $WS_2$ lattice indicates that not much local strain is introduced by the substitution of a W atom in $WS_2$ with Er since the relative size of both species is quite similar. This is evident from the slight bond-length distortions near the Er impurity. For instance, in pristine $WS_2$, we find 2.42 and 3.18 Å for W–S and W–W bond lengths, respectively, while in the Er:$WS_2$ case (with $C_1$ symmetry), Er–S and Er–W are 2.60 and 3.28/3.29 Å. Keeping high symmetry, the atomic configuration of the Er-doped $WS_2$ monolayer remains in the $D_{3h}$ point group, with 2.61 and 3.29 Å for the Er–S and Er–W bond lengths, respectively. Overall, these local distortions remain relatively small (<8% bond-length) across all studied supercell charge states (+1 to -3) for both $C_1$ and $D_{3h}$ cases. We note, however that the lower symmetry structures are 100–300 meV more stable in energy than the high-symmetry structures.

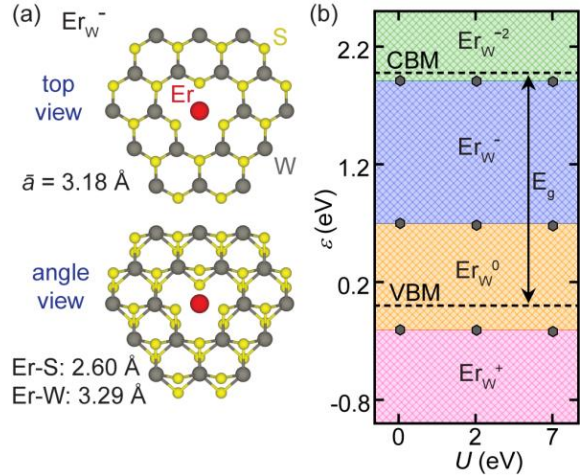

**Figure 1.** Structural and thermodynamic stability of $Er_W$ in monolayer $WS_2$. (a) Atomic structure of the negatively charged defect ($Er_W^-$) in top and angle (20 deg.) views. (b) Charge state transition energies of $Er_W$ within PBE and PBE+U.

In general, defect stabilities are characterized by calculating the formation energies with respect to the Fermi level.[42,43] The difference between formation energies leads to the charge state transition energies,[43] as shown for this system in Figure 1(b). In calculating the formation energies, we employ the correction scheme by Freysoldt et al.[44,45] for charged defects suitable for 2D materials within the supercell approach. First, we find that a substitutional Er in monolayer $WS_2$ is most stable in its neutral ($Er_W^0$) and negatively charged ($Er_W^-$) states. Both charge states are stable across a wide (~0.7 eV) range of the bandgap, suggesting that $Er_W^0$ and $Er_W^-$ could coexist under photoexcitation conditions. We also find $Er_W^{-2}$ to be stable along a narrow range, which suggests it may also be observed experimentally. Due to the high confinement and strong correlations found in heavy atoms[19,46] (e.g., REs), Er(f) states might not be well described by the standard PBE functional within DFT. To preempt this problem, we explore the use of the DFT+$U$ approach,[21] that has been applied to other Er defect systems in the past.[18,23] We calculate the charge state transition levels for different values of the Hubbard $U$ value in Fig. 1, with $J = 1$ eV in all calculations. We find that changes in formation energy show only a slight dependence on $U$ (up to 7 eV), typically of the order of $10^{-2}$ eV. To obtain the $U$ parameters self-consistently, methods based on constrained RPA[47] or linear response[48] could also be used.

We now focus on the electronic properties of $Er_W^-$, starting with the electronic band structure and DOS, shown in Figure 2(a) in the absence of spin-orbit coupling (SOC). For reference, we first note that $Er^{3+}$ has eleven electrons in the 4f manifold. Following Hund's rule, eight of these electrons are paired, leaving three unpaired electrons that give rise to a local magnetic moment. Figure 2(a) shows a similar electronic configuration for substitutional Er in $WS_2$, namely, we find three unoccupied Er(f) states within the bandgap. On the other hand, the highest Er(f) occupied states lie on the same (minority) spin channel and just below the valence band maximum (VBM), which we have set to zero in all band structures and DOS plots. A high spin unbalance, resulting from the partial 4f occupation in Er causes the occupied Er(f) states in the other (majority) spin channel to lie deep within the valence band (VB).

We emphasize the distinction of the total charge state of the supercell and the local charge state of the Er ion. In fact, we find that when changing the overall charge state of the defect, the Er ion remains in its +3-charge state, while the occupancies of localized dangling bonds of W and S atoms are changed instead. As discussed later, these dangling bonds lead to localized defect states reminiscent of a plain W-vacancy ($V_W$), which correlates well with previous studies in TMDs.[49,50] We warn that the relative energy alignment of these states — here alternating with the $Er^{3+}$ levels — must be regarded as tentative given the limitations derived from the exchange-correlation treatment in DFT calculations.[51] On the other hand, we find that the dangling bond states have a strong W(d) character while the empty states of the Er ion stay nearly intact (see DOS plots in Figure 2(a)). These low levels of hybridization suggest the electronic properties of $Er^{3+}$ remain mostly unchanged relative to those found in other host crystals (e.g., YSO).[3]

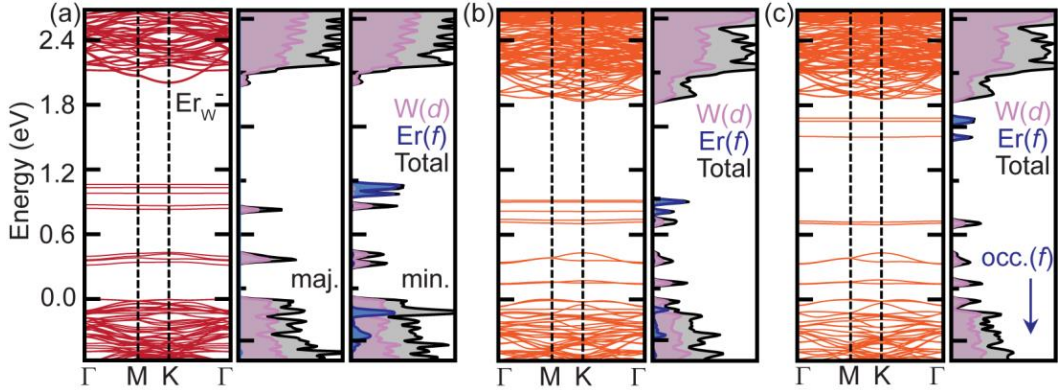

**Figure 2.** Electronic band structure and partial DOS of $\mathrm{Er_W^-}$ in monolayer WS$_2$. We calculate the band structures within DFT, (a) without the inclusion of SOC and (b) with SOC. We also calculate the band structures within DFT+$U$ ($U$ = 2 eV, $J$ = 1 eV shown as an example) including SOC, as shown in (c). DOS labels represent majority and minority spin channels, while pink, blue and gray fills represent partial W($d$), Er($f$) and total DOS contributions, respectively. Red/orange colors distinguish band structure results without/with SOC. For the DFT+$U$ case, the Er($f$) occupied states go deep within the VB (indicated by the blue arrow in (c)).

Because highly aspherical electronic orbitals are present in heavy atoms, SOC is expected to play an important role both for W and Er. The results in Figure 2(b), however, confirm this view only partly: most band features remain unchanged in the presence of SOC, except for a slight reordering of the occupied defect states and a minor decrease in the DOS of Er. On the other hand, inclusion of SOC introduces significant changes in the total energies (of order ~0.06 eV/atom), making the overall system more stable. For this reason, we focus our discussion below on results that do include SOC (unless otherwise stated).

Recalculating this band structure at the DFT+$U$ level, we find that even small $U$ values (2 eV in Figure 2c) lead to a substantial increase of the gap between occupied and unoccupied states, pushing occupied states well within the VB as indicated by the blue arrow in Figure 2(c). The experimental value for the $^4I_{15/2}$–$^4I_{13/2}$ transition in $Er^{3+}$ (being an $Er^{3+}$ HOMO-LUMO transition in our picture) has been determined to be ~0.80 eV and almost independent of the host crystal for a wide range of materials.[3,13,52,53] From our calculated KS energies in the $U = 0$ case, we deduce an energy difference ('transition energy') of 0.91 eV, which surprisingly only slightly overestimates the experimental value. This transition energy, however, grows as the value of $U$ is increased, reaching 3.88 eV for $U = 3$ eV. This finding highlights the limitations of ground-state DFT to study excited state properties for these systems and shows the need for more accurate methods beyond ground state DFT, such as ΔSCF,[54,55] time-dependent DFT (TDDFT),[39] GW/BSE[56] or correlated embedding methods.

In the following, we turn our focus to the optoelectronic features of the $\mathrm{Er_W^-}$ point defect. Typical transition rates in RE ions are comparatively slow, with absorption/emission features that may be surprisingly narrow, but weak in intensity. Thus, we anticipate any absorptive feature in $\mathrm{Er_W^-}$ coming from $Er^{3+}$ to be relatively weak. Turning back to the band structure in Figure 3(a) for $U = 0$, we calculate the projected partial charge densities of the relevant states (Figure 3(b)) to visually inspect the atomic nature of each of these, along with potential hybridization. We see that the Er($f$) HOMO state (close to the VBM) is localized without any appreciable hybridization. Further, what we identify as the defect's HOMO state, appears localized on the Er ion in Figure 3(b) but has mostly a W($d$) character, which is the reason why we characterize it as a defect state (def.).

Next level up, we have a slightly delocalized (defect) state that is identified as the LUMO. Clearly, the overlap between the Er($f$) HOMO state and this LUMO is minimal. Thus, we focus on the empty states that are above this LUMO level, which we find to be localized almost entirely on the Er ion ($f$ states). Here we show only two of them as examples, namely the lowest (LUMO+1) and the highest (LUMO+3) in energy among the three, noting that the energy difference between them is rather small (<90 meV). Again, we see negligible hybridization of these Er($f$) states. Overlap between the Er($f$) HOMO state and these highest empty states within the bandgap should be appreciable, which would correspond to an Er($f$–$f$) transition within monolayer WS$_2$.

To understand the optical features arising from transitions between these electronic states, we turn our focus to the absorption coefficients, as calculated from the imaginary and real components of the dielectric tensor. We expect to see most contributions from the relevant localized defect states in the vicinity (±1 eV) of the Fermi level, so we focus on the low-energy region of the optical spectrum in all plots. For $\mathrm{Er}_\mathrm{W}^-$, a few transitions are observed within this spectral region at the independent particle approximation (IPA) level (lower panel in Figure 3(c)), with the lowest (in-plane) ones centered at 0.39, 0.60 and 0.96 eV. At this level, it is not possible to determine the state origins of each excitation within DFT, but we note that these depend on the energy gap between the KS states. Correlating the energy gaps with each absorption maximum, we conclude that the peaks at ~0.96 eV originate from Er($f$–$f$) transitions. This value is slightly greater than that observed experimentally (~0.80 eV) though this type of overestimation is typical.[38,57] Using the Casida equation, one can expect to obtain more accurate excitation energies, since it is based on linear response TDDFT, which is defined rigorously for electronic excitations.[39,58] The corresponding spectrum within this formalism at the Tamm-Dancoff approximation (TDA) level for $\mathrm{Er}_\mathrm{W}^-$ is shown in the upper panel of Figure 3(c). We now find one main (lowest) transition in the vicinity of 0.9–1.0 eV, with the lower-energy peaks previously observed at the IPA level being absent, and a stronger peak arising at ~1.5 eV.

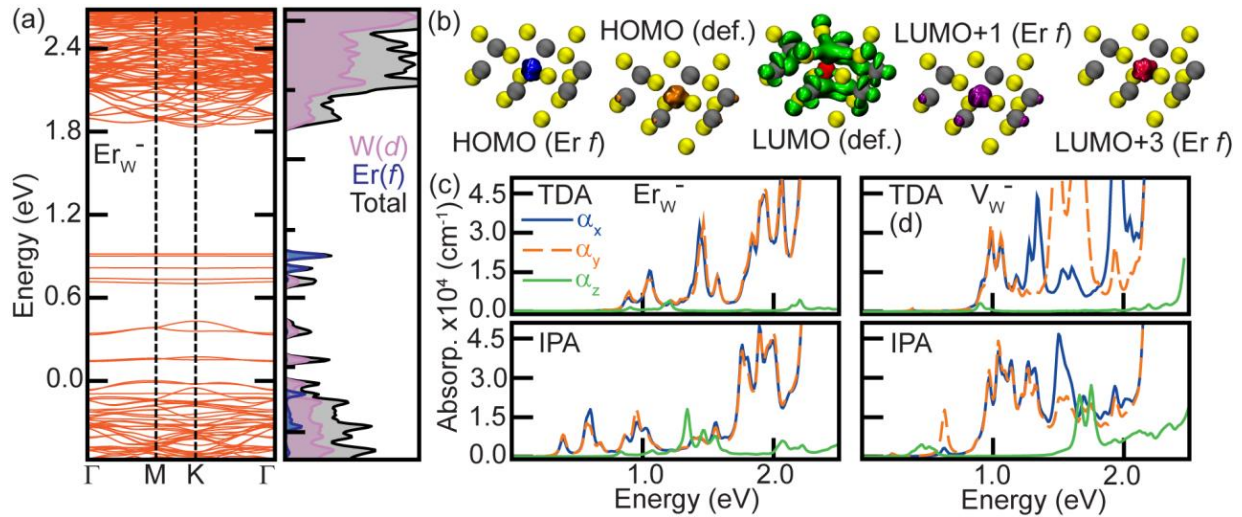

**Figure 3**. Optoelectronic properties of the $\mathrm{Er}_\mathrm{W}^-$ center as compared to the bare negatively charged W-vacancy ($\mathrm{V}_\mathrm{W}^-$) in monolayer $WS_2$. (a) Electronic band structure of $\mathrm{Er}_\mathrm{W}^-$ (same as Figure 2(b)). (b) Partial charge density plots for few relevant states across the vicinity of the Fermi level. 'HOMO (def.)' and 'LUMO (def.)' represent what we identify as defect states inside the bandgap, while 'HOMO (Er $f$)' and 'LUMO+$x$ (Er $f$)', $x$ = 1, 3 are levels defined only with respect to $Er^{3+}$ occupancies. Optical absorption coefficients for (c) $\mathrm{Er}_\mathrm{W}^-$ and (d) $\mathrm{V}_\mathrm{W}^-$ within two levels of approximation, IPA and TDA. Blue solid and orange dashed lines represent in-plane components, while green solid lines represent the out-of-plane components.

To better understand the origin of these peaks, we first inspect more closely the optoelectronic signatures of a bare W-vacancy defect states, as well as the possible changes introduced to these states upon doping. To this end, we perform the same set of optoelectronic calculations for $\mathrm{V}_\mathrm{W}^-$ (Figure 3(d)). Interestingly, we find similar features in both cases, with some slight changes to the relative intensity and shape of the overall spectrum. In principle, the overall differences observed between the optical signatures of the $\mathrm{Er}_\mathrm{W}^-$ and $\mathrm{V}_\mathrm{W}^-$ centers should be attributed to $Er^{3+}$, but to do this rigorously, it is important to disentangle Er and $V_W$ contributions to the peaks identified in Figure 3(c). To do so, we consider the oscillator strengths $f$, wavefunctions overlap, and transition dipole moments (TDM) for three main cases: (*i*) the isolated $Er^{3+}$ ion, (*ii*) the $\mathrm{V}_\mathrm{W}^-$ defect, and (*iii*) the $\mathrm{Er}_\mathrm{W}^-$ center. Comparison between these results allows for an efficient deconvolution of the absorption features into contributions from various excitations, while helping understand the overall optoelectronic properties expected for a $\mathrm{Er}_\mathrm{W}^-$ center in $WS_2$.

Figure 4(a) zeroes in on the case of bare $Er^{3+}$: the top panel shows $f$ as calculated from the Casida equation at the TDA level. We focus only on transitions between the HOMO and the three empty $f$ states (shown in red), as well as those originating from the level right below the HOMO (HOMO-1) and the same three empty $f$ states (shown in black). In the bottom panel, we plot $f$ as calculated from DFT (based on standard overlap of the wavefunctions). As in Figure 3, the results from TDA are slightly blue shifted as compared to those from IPA. Further, TDA level calculation yields a slightly broader transition, which comes from the inclusion of many more states in the construction of the Casida equation, as well as coupled transitions that are considered independent at the DFT level.

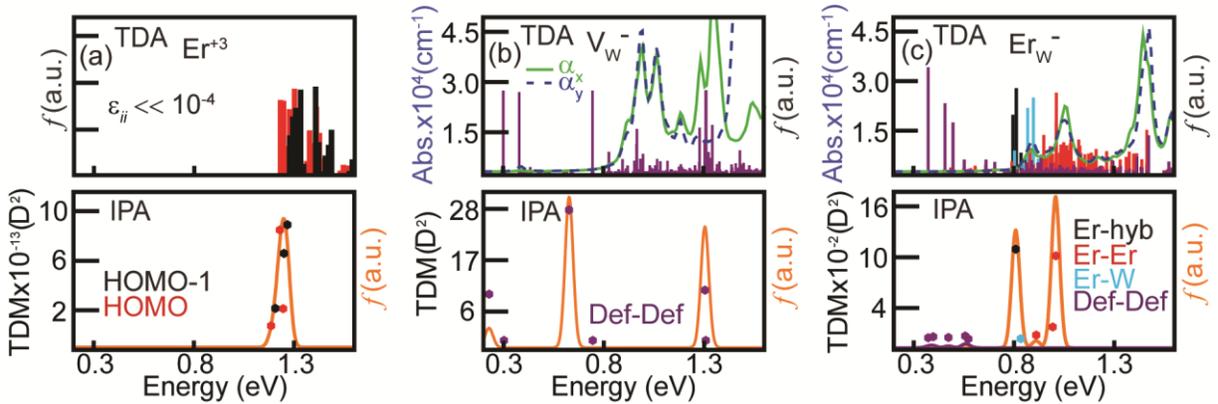

**Figure 4**. Oscillator strengths ($f$) and transition dipole moments (TDM) calculated at the IPA and TDA levels for (a) an isolated $Er^{3+}$ ion, (b) the $V_W^-$ defect, and (c) the $Er_W^-$ center. Solid green and dashed blue lines represent the corresponding in-plane absorption spectra from Figure 3 (note the change in legend). For the isolated $Er^{3+}$, the dielectric components $\varepsilon_{ii}$ lie well below the threshold to be considered as non-zero in VASP (~$10^{-4}$), resulting in the inability to plot the corresponding absorption spectrum within this energy window. Solid orange lines correspond to oscillator strengths at the DFT level calculated from wavefunction overlap. Color-coded scatters/bars represent the different excitation contributions to the optical features.

From the overlap of the DFT wavefunctions, we have access to the transition dipole moments (lower panel in Figure 4(a). Not surprisingly, these TDMs are vanishing, which may be the main reason for the absence of any low-energy absorption features for $Er^{3+}$ at the TDA level within VASP (dielectric tensor components $\varepsilon_{ii}$ lying below the non-zero threshold of $10^{-4}$–$10^{-5}$). A relevant observation in these calculations is the overestimation of the energies corresponding to the $f$–$f$ transitions within $Er^{3+}$ noting that, in principle, any crystal field effect should be only on the order of $10^2$ cm$^{-1}$ for $Er^{3+}$ $f$ states.[13] It is thus natural to expect the transitions to be overestimated for the case of $Er^{3+}$ in $WS_2$, which in fact correlates with what we have calculated.

Let us now turn to the case of the $V_W^-$ center in $WS_2$. For simplicity, we consider only transitions from the defect's HOMO state to all empty defect states within the bandgap. From the top panel of Figure 4(b) we already see relatively good agreement between the most prominent oscillator strengths and the absorption spectrum suggesting the latter arises mainly (though not fully) from contributions associated to the considered transitions. As compared to the bottom panel of Figure 4(b), we again obtain (as expected) what seems to be a blue shift of the transitions at the IPA level. There are two prominent transitions that lie at 0.62 and 1.3 eV, with larger TDMs as compared to the transition for the bare $Er^{3+}$. Importantly, these features correspond to the bare vacancy, and they should change as the Er ion is incorporated owing to, for example, electrostatic interactions introduced by the ion.

In Figure 4(c), the picture is more complex due to mixed excitations to/from defect states and Er($f$) states that need to be taken into account. The color coding in Figure 4(c) captures transitions between $Er^{3+}$ states and slightly hybridized states ('Er-hyb'), between $Er^{3+}(f)$ states ('Er-Er'), between $Er^{3+}$ and the defect's LUMO state ('Er-W'), and between defect states ('Def-Def'). Note that the bright transitions prominent in Figure 4(b) have now shifted towards lower energies (shown in purple in Figure 4(c)) and become substantially less bright. The latter is likely a consequence of changes in the extension and shape of the occupied and unoccupied $V_W^-$ states, impacted differently by the inclusion of Er.

We finally focus on the region around 1.0 eV of the spectrum, where we have obtained a transition with TDA peaks centered at 0.89 and 1.05 eV. We indeed find that most of the excitation character for these optical features comes from Er-Er transitions at both the DFT level and within the Casida approach. Despite the overestimation of optical transitions within DFT, both results suggest that a substitutional Er impurity in monolayer $WS_2$ should be optically active within the telecom band, a response reminiscent of the $^4I_{15/2} \leftrightarrow {}^4I_{13/2}$ transition in $Er^{3+}$.

## IV. CONCLUSIONS

DFT calculations indicate that an Er impurity occupying a W site in a $WS_2$ monolayer forms a stable point defect. The Er impurity introduces only a minor distortion in the $WS_2$ lattice, a manifestation of the close match between the Er and W atomic radii. Formation energies favor the neutral and negatively charged states, but other charge states (e.g., $Er_W^{-2}$) may be experimentally observable. Valence electrons from the Er atom (along with any excess electrons in the super-cell) localize at the dangling bonds from nearby sites, hence leaving behind an $Er^{3+}$ ion core. The system's ground state energy lies slightly below the valence band maximum, and there is some spatial overlap between the defect and the $Er^{3+}$ ground state orbitals although the level of hybridization is low. The latter also applies to the set of excited states some of which remain confined to the Er site (and can be associated to the excited $Er^{3+}$ orbitals) while the rest delocalize (and can be loosely seen as $V_W^-$ states).

We built on the strong differences between these two classes of states to characterize the optical response of the defect, which we determined through optical absorption and oscillator strength calculations. In particular, we found Er (*f-f*) absorption near 0.9 eV, hence suggesting the $^4I_{15/2} \leftrightarrow ^4I_{13/2}$ transition in $Er^{3+}$ remains active in the 2D host. We also want to emphasize the possible limitations of DFT to describe strongly correlated systems such as the Er ion and systems of multi-reference character. While the DFT+*U* approach can overcome some of those shortcomings in particular for ground state calculations, more accurate methods, such as correlated embedding methods[17–20] are necessary to predict accurate excitation energies. For such methods, our calculations can serve as the starting point. Future work will include comparison of our findings with these methods as well as the study of the role of phonons in these optical transitions.

All in all, these findings point to Er:WS$_2$ as a potentially promising material platform for applications that rely on single photon emitters integrated to photonic structures. Whether or not the $Er^{3+}$ retains the useful spin and magneto-optical properties observed in other crystal hosts[59] is a most intriguing question, whose answer will require, nonetheless, additional computational and experimental work.

After initial submission of this manuscript, we became aware that electronic structure calculations for the electronic and optical properties of Er:WS2 using molecular orbital theory within the SIESTA software package have been published.[60] Besides differences between these two studies in terms of methods for excited-state properties (molecular orbital theory/IPA and Casida equation) and software packages (SIESTA/VASP) we find qualitative agreement on the main findings.

**ACKNOWLEDGEMENTS**

G.I.L.M. acknowledges funding from the NSF CREST IDEALS, through grant number NSF-HRD-1547830. V.M.M. and C.A.M acknowledge support from the National Science Foundation under grant NSF-ECCS-1906096. All calculations were performed using the computational facilities of the Flatiron Institute. The Flatiron Institute is a division of the Simons Foundation.